\documentclass[twocolumn,showpacs,preprintnumbers,amsmath,amssymb]{revtex4}

\usepackage{graphicx}
\usepackage{dcolumn}
\usepackage{bm}

\begin{document}


\title{Conductance fluctuations in the presence
of spin scattering}

\author{Jun-ichiro Ohe}
\author{Masayuki Yamamoto}%
\author{Tomi Ohtsuki}%
\affiliation{%
Department of Physics, Sophia University, Kioi-cho 7-1, Chiyoda-ku, Tokyo 102-8554, Japan}


\date{\today}


\begin{abstract}
Electron transport through disordered systems that include
spin scatterers is studied numerically.
We consider three kinds of magnetic impurities:
the Ising, the XY and the Heisenberg.
By extending the transfer matrix method to include the spin
degree of freedom,  the two terminal conductance is calculated.
The variance of conductance is halved as
the number of spin components of the magnetic impurities increases.
Application of the Zeeman field in the lead causes
a further halving of the variance under certain conditions.
\end{abstract}

\pacs{73.23.-b, 72.25.Rb, 73.63.Nm}
\maketitle

\section{Introduction}

Quantum transport phenomena that involve the carrier's spin degree of freedom
have attracted a lot of attention during the past decade.\cite{Datta,Bruno}
A number of studies have analyzed the spin polarized transport
in ballistic regime and reported intriguing phenomena,\cite{Koga,Matsuyama}
and have opened up the possibility of new spintronic devices.
It is also interesting to study the spin-dependent transport in
diffusive and chaotic regimes because the interference of
coherent electron waves shows a number of characteristic effects
when the spin degree of freedom is taken into consideration.
One of the special characteristics of diffusive and chaotic
systems is the fluctuating nature of transport coefficients such as 
conductance.\cite{Lee}
It is well understood that such fluctuations do not depend on
the details of the sample parameters, but
depend only on the symmetry of the system.\cite{Lee,Beenakker}
The two relevant symmetries here are time-reversal symmetry (TRS)
and spin-rotation symmetry (SRS).
TRS is broken by applied magnetic fields or
by magnetic scattering due to magnetic impurities
or magnetic domain walls.
If TRS is broken, systems are classified as unitary,
regardless of whether or not SRS is broken.
The spin-orbit interaction breaks SRS but preserves
TRS, and in this case the systems are classified as symplectic.

When the spin degree of freedom is taken into account, the description
of the conductance fluctuations becomes more complex.
Conductance fluctuations of 2-dimensional systems which are coupled to
the Ising spin glass is reported.\cite{Cieplak}
It has been reported that the reduction of variance of
conductance takes place
due to the Zeeman splitting in the sample region.
\cite{Altshuler,Debray,Birge}
Altshuler and Shklovskii had shown that
the variance of the conductance is described by\cite{Altshuler}
\begin{eqnarray}
<{\delta G}^2>=s^2\frac{3K}{\beta}(\frac{e^2}{\hbar\pi^3})^2b_d,
\end{eqnarray}
where $b_d$ is a dimension-dependent factor that is of the order of
unity.
$\beta$ is equal to 1,2 and 4 for orthogonal, unitary and symplectic
systems, respectively.
The quantity $K$ is equal to the number of noninteracting series of
energy levels with $s$-fold spin degeneracy.

Recent works have pointed out that the universal conductance fluctuations
(UCF) in a chaotic quantum dot in the presence of spin-orbit scattering shows
new features.\cite{Folk,Halperin,Aleiner}
They have shown that the UCF in the presence of spin dependent
scattering is interesting not only from the theoretical point of
view but also from the experimental view point.

The transport properties of mesoscopic systems 
depend not only on the sample but also
on the states in the leads through which currents flow into
and out of the sample and through which voltages are
measured.\cite{buettiker88}
How the transport properties and
the universality class are changed
by modulating lead states is a very interesting question,
especially when the spin degree of freedom plays a role.

In this paper, we investigate the influence of the spin
scattering on transport properties in disordered systems.
We consider magnetic impurities in sample
region and the Zeeman field in the lead.
Three types of magnetic impurities are considered.
We call these Ising, XY and Heisenberg, depending on the
number of spin components.
In order to calculate two terminal conductance, we
employ the transfer matrix method\cite{Pendry}
that is extended to include the spin degree of freedom.
Magnetic impurities remove the spin degeneracy and break
TRS in certain cases.

We find that the variance of the conductance is halved as the
number of spin components of the magnetic impurities increases. 
When the Zeeman field is applied in a lead,
a further reduction of the variance is observed.
In order to observe the crossover of the universality class
with the increase of the Zeeman field in the lead,
we study the level spacing distribution of transmission eigenvalues.
Part of this work has been presented in the international conference,
^^ ^^ Localisation 2002''.\cite{oyos,yos}

\section{Magnetic scattering}

We consider a two dimensional (2D) system connected to two electrodes.
The 2D system is constructed in the $x$- and $y$- directions 
and the current flows in the $x$ direction.
There is an exchange interaction between the electron spin and the static
local spins in the system.
The one-electron Hamiltonian is
\begin{eqnarray}
H &=& H_0+H' ,\\
H_0 &=& \sum_{i,\sigma}W_{i}c_{i,\sigma}^+c_{i,\sigma}
- \sum_{<i,j>,\sigma,\sigma'} V_{i \sigma,j\sigma'} 
c_{i,\sigma}^{\dagger}c_{j,\sigma'},\\
H' &=& -J\sum_{i,\sigma,\sigma'}c_{i,\sigma}^\dagger
\boldsymbol{\sigma}_{\sigma,\sigma'}c_{i,\sigma'}\cdot {\bf S}_i,
\end{eqnarray}
where $c_{i,\sigma}^\dagger (c_{i,\sigma})$ denotes the creation
(annihilation) operator
of electron at the site $i$ with spin 
$\sigma(=\pm 1)$ on the 2D square lattice.
On-site energy $W_i$ denotes the non-magnetic random potential distributed
independently and uniformly
in the range $[-W/2,W/2]$.
The hopping is restricted to nearest neighbors.

\begin{figure}[ttt]
\begin{center}
\includegraphics[scale=0.75]{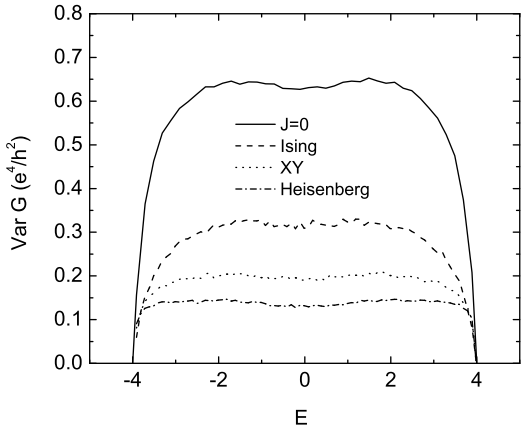}
\caption{Variance of conductance in the presence of
magnetic impurities. System size is $30\times 30$. $W$ is set to be $4.0$
and $J=0.4$.
}
\end{center}
\end{figure}

\begin{figure}[ttt]
\begin{center}
\includegraphics[scale=0.75]{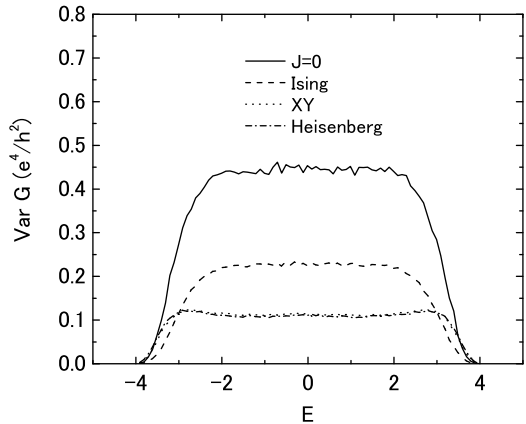}
\caption{Variance of conductance when the sample contains both the
magnetic field and
magnetic impurities. Parameters are the same as in Fig.~1.}
\end{center}
\end{figure}

We investigate how the variance of the conductance distribution
 of $H_0$
changes due to the presence of $H'$.
The variance of conductance distribution for
$H_0$ is determined by the symmetry, which is
controlled by the hopping term,
$V_{i \sigma,j\sigma'}$.
If it is set to $2\times2$ unit matrix, $H_0$ belongs
to the orthogonal class.
$H_0$ belongs to the unitary class when
\begin{equation}
V_{i,i+\hat{y}} =
\left( 
\begin{array}{cc}
\exp(i\phi) & 0 \\
0 & \exp(i\phi) \\ 
\end{array}
\right),
\end{equation}
where $\phi$ is a Peierls phase which is distributed uniformly
in the range $[0,2\pi)$.
To realize systems  belonging to the symplectic class,\cite{ando}
we set the hopping 
\begin{equation}
V_{i,i+\hat{x}} =
\left( 
\begin{array}{cc}
\cos \theta & \sin \theta \\
-\sin \theta & \cos \theta \\ 
\end{array}
\right)
\end{equation}
and
\begin{equation}
V_{i,i+\hat{y}} =
\left( 
\begin{array}{cc}
\cos \theta & i \sin \theta \\
i \sin \theta & \cos \theta \\ 
\end{array}
\right).
\end{equation}
where the parameter $\theta$ denotes the strength of the
spin-orbit interaction.

The additional term $H'$ is the spin scattering term.
$\boldsymbol{\sigma}$ is the Pauli spin matrix and ${\bf S}_i$ is the
static local spin.
We consider three types of magnetic impurities: Ising, XY and 
Heisenberg where
${\bf S}_i=(0,0,S_z)$, $(S_x,S_y,0)$ and $(S_x,S_y,S_z)$, respectively.
The absolute value of ${\bf S}_i$ is set to unity,
and their components are distributed randomly according to the
uniform distribution on $n$-dimensional
sphere, $n$(=1,2 or 3) being the number of spin components.

In order to calculate the conductance, we employ the transfer matrix method
\cite{Pendry}
and extend it to include the spin degree of freedom.
The conductance $G$ is given by the Landauer
formula\cite{landauer57,landauer75} as
\begin{eqnarray}
G=\frac{e^2}{h}{\rm tr}(tt^\dagger)=\frac{e^2}{h}\sum_i\tau_i
\end{eqnarray}
where $t$ is the transmission matrix including the spin degree of
freedom and $\tau_i$ is the transmission eigenvalue.
In the present simulation, the system size is $30\times 30$
in units of the lattice spacing

\begin{figure}[ttt]
\begin{center}
\includegraphics[scale=0.75]{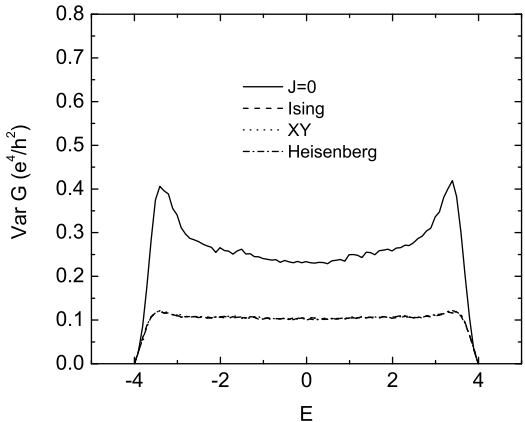}
\caption{Variance of conductance when the sample contains both the
spin-orbit interaction and
magnetic impurities. Parameters are the same as in Fig.~1.}
\end{center}
\end{figure}

In Fig.1 we show the variance of conductance in the presence of
magnetic impurities where $H_0$ belongs to the orthogonal class.
$W$ is set to be 4.0 and $J=0.4$.
$10^4$ ensemble averages are taken for each data.
When $J=0$,
there is no magnetic scattering and the variance
is close to that expected for
the universal conductance fluctuation of orthogonal systems.\cite{Lee}
For convenience, let us denote the variance of
$\sum_i{\tau_i}$ for the conventional orthogonal
class as $V_{\rm orth}$ and $G/(e^2/h)=\tilde{G}$, so that
in the case of $J=0$,
${\rm Var } \tilde{G}= V_{\rm orth}$.
$V_{\rm orth}$ is 1/2 for the chaotic cavity,
8/15 for the quasi-1D wire, 0.74 for 2D, and 1.2 for 3D
systems, respectively.\cite{Beenakker,Lee}

In the presence of the Ising type impurities, no spin flip
processes occur as in the conventional non-magnetic system.
However, the up spin and down spin electrons are
described by different wave functions, because the exchange field of
magnetic impurities lifts the spin degeneracy.
With sufficiently large $J$, these wave functions are no longer
correlated, and the variance is represented by
\begin{equation}
{\rm Var } \tilde{G}={\rm Var } 
\tilde{G}_{\uparrow }+{\rm Var } \tilde{G}_{\downarrow },
\end{equation}
where $\tilde{G}_{\uparrow (\downarrow) }$ is the conductance
through the up (down) spin channel.
Since both of the variances  $\tilde{G}_{\uparrow }$ and
$\tilde{G}_{\downarrow }$ are $V_{\rm orth}/4$, the sum
$\tilde{G}$ becomes $V_{\rm orth}/2$

While Ising type impurities do not rotate the spin direction,
the spin flip process occurs in XY type impurities.
Then the variance is simply given by ${\rm Var} \tilde{G}=V_{\rm orth}/4$,
since the factor of 4 coming from the spin degeneracy in the
conventional non-magnetic orthogonal class is missing.
Though the Hamiltonian is complex due to $\sigma_y$,
the statistics of the transmission eigenvalues
as well as the energy level statistics are that of
the orthogonal class.  This can be seen if
we define the time-reversal operator $T$ as
$T=i\sigma_x K$, where $K$ denotes the
complex conjugation.
This operator is anti-unitary, satisfies $T^2=1$ and
commutes with the Hamiltonian
of the system including XY type impurities.

For Heisenberg and XY type impurities, spin flips occur.
However,  when $H'$ includes the Heisenberg type scatterers,
the Hamiltonian no longer commutes with the time-reversal operator, and
the system is classified into the unitary class.
Therefore, the variance
is further reduced by a factor of 2.

From these results, the variance of conductance
in the presence of impurities is given by
\begin{eqnarray}
{\rm Var} \tilde{G}=\frac{V_{\rm orth}}{2^{n}},
\end{eqnarray}
We have numerically inverstigated a square 2D system, but
the argument above is general, and this relation should be valid in
other dimensions.\cite{oyos}

Figure 2 shows the change of the variance when $H_0$
is in the unitary class.
Parameters are the same as in  Fig.~1.
We consider the random magnetic field and the phase in the transfer
integral is given by Eq.~(2).
Without magnetic impurities, the system belongs to the unitary class
with spin degeneracy and the variance is $V_{\rm orth}/2$.
Ising type magnetic impurities removes the spin degeneracy
and the variance becomes $V_{\rm orth}/4$.
Spin flips occur due to the XY type magnetic impurities and
the variance becomes half of the Ising case.
As shown in  Fig.~2, the variance of the conductance when $H'$ includes
the Heisenberg type impurities is the same as in the case of the XY type,
because the Hamiltonian of neither system
commutes with the time reversal operator.

The reduction of the variance is also
obtained when $H_0$ includes the spin-orbit interaction and the
Hamiltonian belongs to the symplectic class (Fig.~3).
Once the magnetic impurities are included, irrespectively of the
type of the magnetic scatterers,
the variance becomes $V_{\rm orth}/8$ from $V_{\rm orth}/4$.

\begin{figure}[ttt]
\begin{center}
\includegraphics[scale=0.75]{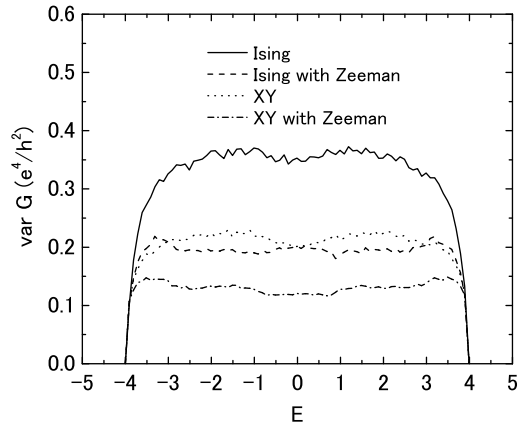}
\caption{The effect of the Zeeman coupling in a lead on the variance.
$H_0$ belongs to the orthogonal class, $W=3.0$ and $Z=3.0$}
\end{center}
\end{figure}

\section{Effect of leads}

\begin{figure}[ttt]
\begin{center}
\includegraphics[scale=0.75]{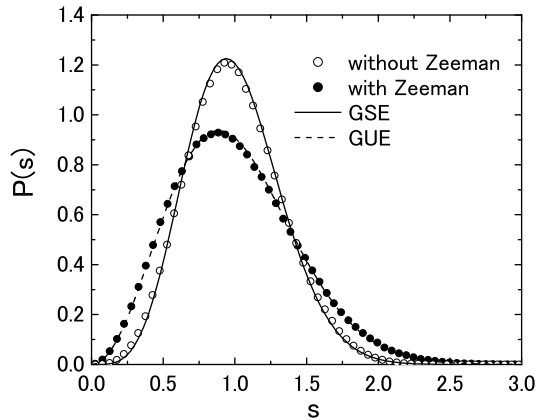}
\caption{Spacing distribution of $\tau$ for the sample with the spin-orbit
interaction.
We set $W=2.0$ and $Z=3.0$.
The distribution without Zeeman splitting in a lead ($\circ$) fits to
the form of GSE while that with Zeeman ($\bullet$) fits to GUE.}
\end{center}
\end{figure}

We then address 
the question of whether or not transport properties and the universality 
class can change as a result of an asymmetric spin population in the leads,
i.e., when the number of up and down spin channels in the leads
becomes asymmetric as a result of Zeeman splitting.
We show that even if the sample is unchanged, the
universality class can change under certain conditions.

We consider the Zeeman field in one of the leads, setting the other
lead Zeeman field free.
The transverse energy of the channel $(i,\sigma)$ in the lead 
$\varepsilon_{i}^\sigma$ is given by
\begin{equation}
\varepsilon_{i}^\sigma = -2\cos(\frac{i\pi}{L+1})-Z\sigma
\hspace{1cm} (i=1,2,\cdots,L)
\end{equation}
where $L$ denotes the number of sites in the transverse direction
($y$-direction), and
$Z$ denotes the strength of Zeeman splitting and
$\sigma (=\pm 1)$ is the spin index. 
The channel $(i,\sigma)$ is a propagating mode if 
\mbox{$|E_{F} - \varepsilon_{i}^\sigma| < 2.0$},
$E_{F}$ being the Fermi energy.
For example, when we set $E_{F}=-1.1$ and $Z=1.0$,
the number of up (down) spin channels becomes 27 (15) for the sample of
width 30 sites.

Figure 4 shows the variance in the presence of Ising and XY type
impurities in a system including
Zeeman splitting in a lead.
In this simulation, 
we set $W=3.0$, $\theta =\pi/4$ and $E_{F} = -1.1$.
The system size is again set to be $30 \times 30$ in units of the lattice
spacing.
The population of up and down spins in one lead is always set to be
symmetric ($Z=0$), and that in the other lead is varied ($Z=0,3.0$).

From this figure, we observe that the
variance becomes almost half due to the Zeeman splitting in the leads,
even if the sample region is not changed at all.
The effect of Zeeman splitting is similar to that
of increasing the number of spin component of the magnetic impurities.
It should be noted that if the direction of the Zeeman field
is the same as one of the spin components of the scatters,
the reduction in not observed.
For example, in the case of Ising type scatterers,
applying the Zeeman field in the $z$-direction does not change the
variance while the halving of the variance is observed
if we apply the Zeeman field in $x$- or $y$-direction.
The same is true for the case of the XY type scatters.
We need to apply the Zeeman field in the $z$-direction to observe
the halving of the variance.

We then show that the Zeeman field in the lead  changes the
universality class of the system.
To detect the crossover of the universality classes, we investigate 
the spacing distribution $P(s)$ where $s$ is
the interval between neighboring transmission
eigenvalues $\tau$'s.
An ensemble of about $10^6$ samples is simulated
to get good statistics.

\begin{figure}[ttt]
\begin{center}
\vspace{-4cm}
\includegraphics[scale=0.4]{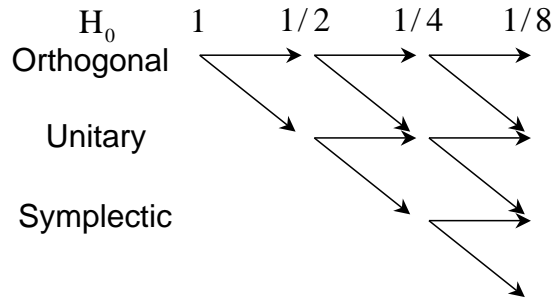}
\vspace{-1.5cm}
\caption{Schematic diagram of the variance of the conductance.
The horizontal arrow indicates the increase of the spin component
of the magnetic scatterers, while the slanted arrow indicates the
addition of the Zeeman field.
The Zeeman field is assumed to contain a component different from that
of the magnetic scatterers.}
\end{center}
\end{figure}

Figure 5 shows the spacing distribution $P(s)$
of the transmission eigenvalue $\tau$
for the sample with spin-orbit interaction.
In the absence of Zeeman splitting in the leads ($\circ$),
$P(s)$ is close to the Wigner surmise for the
Gaussian symplectic ensemble (GSE).
On the other hand, with Zeeman splitting in the leads
($\bullet$), $P(s)$ is close to that for
the Gaussian unitary ensemble(GUE).
The reason for this crossover is that the
asymmetry of up and down spins in the lead destroys 
the self-dual property
of the scattering matrix.\cite{Beenakker}
To be specific, we consider the $(i,j)$ component of the S-matrix
$S_{i,j}$ which describes the transmission from $(i\uparrow,i\downarrow)$
to $(j\uparrow,j\downarrow)$.
The self-duality requires
\begin{equation}
S_{i,j}=\left(
\begin{array}{cc}
a & b\\
c& d
\end{array}
\right)\, ,\,
S_{j,i}
=\left(
\begin{array}{cc}
d & -b\\
-c& a
\end{array}
\right)
\end{equation}
which means that the transmission probability from $i\uparrow$ to
$j\uparrow$ is the same as that from $j\downarrow$ to $i\downarrow$.
This symmetry is broken when we apply the Zeeman field in the lead,
which lifts the degeneracy of $i\uparrow$ and $i\downarrow$.


\section{Summary}

\begin{table*}[ttt]
\caption{Variance of the system in the presence of magnetic impurities
and the Zeeman field.}
\begin{ruledtabular}
\begin{tabular}{lccr}
Universality class of $H_0$ & Additional spin scattering
&Zeeman field in a lead&Variance/$V_{\rm orth}$\\
\hline
Orthogonal&0 & no & $1$\\
&0 & yes & $1/2$\\
&Ising &  no & $1/2$\\
&Ising & yes & $1/4$\\
&XY & no & $1/4$\\
&XY & yes & $1/8$\\
&Heisenberg & irrelevant & $1/8$\\
\hline
Unitary& 0&no& $1/2$ \\
& 0&yes&$1/4$\\
&Ising&no&$1/4$\\
&Ising&yes&$1/8$\\
& XY&irrelevant&$1/8$\\
& Heisenberg&irrelevant&$1/8$\\
\hline
Symplectic&0 & no & $1/4$\\
&0 & yes & $1/8$\\
&Ising & irrelevant & $1/8$\\
&XY & irrelevant & $1/8$\\
&Heisenberg & irrelevant & $1/8$\\
\end{tabular}
\end{ruledtabular}
\end{table*}

We have studied the effect of spin scattering in disordered systems
on the fluctuating nature of the conductance.
We have considered magnetic impurities in a sample and 
calculated transport properties.
Our results show that
the variance of conductance is halved as 
the number of spin components of the magnetic impurities increases.
Halving of the variance of the conductance
is also obtained when the sample includes magnetic field or
spin-orbit interaction.

We have also investigated the effect of the Zeeman splitting in a lead.
Halving of the variance of conductance is obtained when
the direction of the Zeeman field contains the component different
from the component(s) of the magnetic scatterers.
This behavior is reminiscent of the change of the variance in the
superconducting-normal junction.\cite{Beenakker,te92,mbj93}
Analyzing the transmission eigenvalues,
the universality class has been shown to be
changed  by the Zeeman field in the lead.
The results are summarized in Table I and schematically shown in Fig.~6.

Before concluding, we relate our results with that of Aleiner and
Fal'ko.\cite{Aleiner}
They have obtained for the chaotic system
\begin{equation}
{\rm Var}\tilde{G}=\frac{s}{4\beta\Sigma},
\end{equation}
where $s=1,2$ indicates the Kramers degeneracy and $\Sigma=2$ if
the spin flip process is present and $\Sigma=1$ otherwise.
$\beta=$1,2 or 4 is determined by the universality class.
Setting $V_{\rm orth}=1/2$ we recover their interesting result.
For example, when $H_0$ is classified into the orthogonal class,
and $H'$ includes the XY type magnetic impurities,
$\beta=1, s=1$ and $\Sigma =2$, which gives
${\rm Var}\tilde{G}=\frac{1}{8}=V_{\rm orth}/4$. 
Therefore, the present results are the extension of Aleiner and
Fal'ko \cite{Aleiner} to higher dimensions and to the inclusion
of the effect of the Zeeman splitting in the lead.

\begin{acknowledgments}
The authors are grateful to B. Kramer and S. Kettemann
for valuable discussions and K. Slevin for fruitful discussions and
the critical reading of the manuscript.
One of the authors (J.O.) was supported by Research Fellowships of
the Japan Society for the Promotion of Science for Young Scientists.
\end{acknowledgments}


\begin{thebibliography}{99}

\bibitem{Datta} S. Datta and B. Das,
Appl. Phys. Lett. {\bf 56}, 665 (1990).
\bibitem{Bruno} P. Bruno, Phys. Rev. Lett. {\bf 83}, 2425 (1999). 
\bibitem{Koga} T. Koga, J. Nitta, H. Takayanagi and S. Datta,
Phys. Rev. Lett. {\bf 88}, 126601 (2002). 
\bibitem{Matsuyama} T. Matsuyama, C. -M. Hu, D. Grundler, G. Meier
and U. Merkt,
Phys. Rev. B {\bf 65}, 155322 (2002). 
\bibitem{Lee} P.A. Lee, A.D. Stone and H. Fukuyama,
Phys. Rev. B {\bf 35}, 1039 (1987).
\bibitem{Beenakker} C. W. J. Beenakker, Rev. Mod. Phys. {\bf 69}, 731 (1997).
\bibitem{Cieplak} M. Cieplak, B. R. Bulka and T. Dietl,
Phys. Rev. B {\bf 44}, 12337 (1991),
Phys. Rev. B {\bf 51}, 8939 (1995).
\bibitem{Altshuler} B. L. Altshuler and B. I. Shklovskii,
Zh. Eksp. Teor. Fiz. {\bf 91}, 220 (1986)
[Sov. Phys. JETP {\bf 64}, 127 (1986)].
\bibitem{Debray} P. Debray, J. -L. Pichard, J. Vicente and P. N. Tung,
Phys. Rev. Lett. {\bf 63}, 2264 (1989).
\bibitem{Birge} N. O. Birge, B. Golding and W. H. Haemmerle
Phys. Rev. Lett. {\bf 62}, 195 (1989).
\bibitem{Folk} J. A. Folk, S. R. Patel, K. M. Birnbaum,
C. M. Marcus, C. I. Duru\"{o}z and J. S. Harris, Jr.,
Phys. Rev. Lett. {\bf 86}, 2102 (2001).
\bibitem{Halperin} B. I. Halperin, A. Stern, Y. Oreg,
J. N. H. J. Cremers, J. A. Folk and C. M. Marcus,
Phys. Rev. Lett. {\bf 86}, 2106 (2001).
\bibitem{Aleiner} I. L. Aleiner and V. I. Fal'ko,
Phys. Rev. Lett. {\bf 87}, 256801 (2001),
Phys. Rev. Lett. {\bf 89}, 79902 (2002).
\bibitem{buettiker88}
M. B\"uttiker, Phys. Rev. B {\bf 38}, 9375, 12724 (1988).
\bibitem{Pendry} J. B. Pendry, A. MacKinnon and P. J. Roberts,
Proc. R. Soc. A {\bf 437}, 67 (1992).
\bibitem{oyos} J. Ohe, M. Yamamoto, T. Ohtsuki and K. Slevin,
Suppl. J. Phys. Soc. Jpn. {\bf 72}, 209 (2003)
\bibitem{yos} M. Yamamoto, T. Ohtsuki and K. Slevin,
Suppl. J. Phys. Soc. Jpn. {\bf 72}, 155 (2003)
\bibitem{ando} T. Ando, 
Phys. Rev. B {\bf 40}, 5325 (1989). 
\bibitem{landauer57}
R. Landauer,
IBM J. Res. Dev. {\bf 1},223 (1957)
\bibitem{landauer75}
R. Landauer, Z. Phys. B {\bf 21}, 247 (1975)
\bibitem{te92}
Y. Takane and H. Ebisawa, J. Phys. Soc. Jpn. {\bf 61}, 2858 (1992)
\bibitem{mbj93}
I.K. Marmorkos, C.W.J. Beenakker and R.A. Jalabert,
Phys. Rev. B {\bf 48}, 2811 (1993)

\end{thebibliography}
\end{document}